\documentclass[apj]{emulateapj}
  \usepackage{apjfonts} 
  \usepackage[usenames,dvipsnames]{xcolor} 
  \usepackage{slantsc}
  \usepackage{amsmath} 
  \usepackage[T1]{fontenc} \definecolor{blue}{rgb}{0,0,1} 
  \usepackage[hyperfootnotes=false]{hyperref}
  \usepackage{CJK} 
  \usepackage{color} 
  \usepackage{graphicx} \hypersetup{colorlinks, citecolor=blue, linkcolor=blue, urlcolor=blue}
  \usepackage{xhfill}

  \newcommand{\Msun}{\,$\rm{M}_\odot$} 
  \newcommand{\lacompass}{\textit{LA-COMPASS }}
  \newcommand{\radmc}{\textit{RADMC-3D }} 
  \newcommand{\co}{$^{12}\textrm{CO}$ } 
  \newcommand{\cco}{$^{13}\textrm{CO}$ }
  \newcommand{\coo}{$\textrm{C}^{18}\textrm{O}$ } 
  \newcommand{\hco}{HCO$^+$ } 
  \newcommand{\twod}{two-dimensional }
  \newcommand{\ditto}[1][.4pt]{\xrfill{#1}~\textquotedbl~\xrfill{#1}}
  \newcommand{\NA}{---}
  
\slugcomment{}

\shorttitle{Modelling Dust and Gas Emission of HD 163296 Disk with Planet--disk Interaction} \shortauthors{Liu et al.}

\begin{document} 
\begin{CJK*}{UTF8}{gbsn} 
\title{New constraints on turbulence and embedded planet mass in the HD 163296 Disk from Planet--Disk Hydrodynamic Simulations} 
\author{Shang-Fei Liu (刘尚飞) \altaffilmark{1,2}, Sheng Jin (晋升)
\altaffilmark{2,3}, Shengtai Li (李胜台)  \altaffilmark{2}, Andrea Isella  \altaffilmark{1} and Hui Li (李晖) \altaffilmark{2}}
\affil{$^1$ Department of Physics and Astronomy, Rice University, 6100 Main St., Houston, TX 77005, USA; shangfei.liu@gmail.com;
isella@rice.edu \\ $^2$ Theoretical Division, Los Alamos National Laboratory, Los Alamos, NM 87545, USA; sli@lanl.gov;
hli@lanl.gov \\ $^3$ Key Laboratory of Planetary Sciences, Purple Mountain Observatory, Chinese Academy of Sciences, Nanjing
210008, China; shengjin@pmo.ac.cn }

\submitted{}

\begin{abstract} 
  { 
    \noindent Recent Atacama Large Millimeter and Submillimeter Array (ALMA) observations of the protoplanetary
disk around the Herbig Ae star HD 163296 revealed three depleted dust gaps at 60, 100 and 160 au in the 1.3 mm continuum 
as well as CO depletion in the middle and outer dust gaps. However, no CO depletion was found in the inner dust gap. 
To examine the planet--disk interaction model, we present results of two-dimensional two fluid (gas + dust) hydrodynamic 
simulations coupled with three-dimensional radiative transfer simulations. In order to fit the high gas-to-dust ratio of 
the first gap, we find the Shakura--Sunyaev viscosity parameter $\alpha$ must be very small ($\lesssim 10^{-4}$) in the 
inner disk. On the other hand, a relatively large $\alpha$ ($\sim 7.5\times 10^{-3}$) is required to reproduce 
the dust surface density in the outer disk. We interpret the variation of $\alpha$ as an indicator of the transition 
from an inner dead zone to the outer magnetorotational instability (MRI) active zone. Within $\sim 100$ au, the HD 163296 
disk's ionization level is low, and non-ideal magnetohydrodynamic (MHD) effects could suppress the MRI, so the disk can 
be largely laminar. The disk's ionization level gradually increases toward larger radii, and the outermost disk 
($r > 300$ au) becomes turbulent due to MRI. Under this condition, we find that the observed dust continuum and 
CO gas line emissions can be reasonably fit by three half-Jovian-mass planets (0.46, 0.46 and 0.58 $M_\textrm{J}$) 
at 59, 105 and 160 au, respectively.
} 
\end{abstract}


\section{Introduction} Thanks to ALMA's unprecedented angular resolution, fine structures of circumstellar disks around nearby
young stars such as HL Tau \citep{2015ApJ...808L...3A} and TW Hya \citep{2016ApJ...820L..40A} are unveiled now. In particular,
the pattern that consists of a series of concentric bright and dark rings has been discovered in the dust continuum emission in
many systems. However, no consensus has been reached on the formation mechanism of the observed ringed structures. One popular
explanation attributes the observed dust gaps to planets embedded in a protoplanetary disk \citep[][hereafter
J16]{2015MNRAS.453L..73D,2015ApJ...809...93D,2016ApJ...818...76J}. Alternatively, zonal flows \citep[e.g,
][]{2009ApJ...697.1269J,2015A&A...574A..68F}, Rossby wave instability \citep{1999ApJ...513..805L, 2000ApJ...533.1023L,
2001ApJ...551..874L}, rapid pebble growth near the volatile icelines \citep[e.g., ][]{2013A&A...552A.137R,
2015ApJ...806L...7Z,2016ApJ...818L..16Z} have been suggested to account for the ringed structures as well. The caveat is that
those ringed features reflect the emission from dust instead of gas.

\citet{2016ApJ...820L..25Y} identified two gas gaps at 32 and 69 au in the HL Tau disk by performing azimuthal averaging on the
\hco image cube to enhance the signal-to-noise ratio and measure the radial profile of \hco integrated intensity. The inner gas
gap is coincident with the dust continuum gap, while the outer gap is located at the bright continuum ring. It is not clear
whether the outer \hco gap can be related to planet--disk interaction or other scenarios without involving a planet, e.g, CO
snow line. On the other hand, \citet{2016MNRAS.459L...1D} suggested that the observed dust gaps of the HL Tau disk may not
reflect the obscured gas gaps. They performed the smoothed particle hydrodynamics (SPH) simulations to demonstrate that embedded
low-mass planets ($\sim 0.1 M_{\rm J}$) could expel millimeter-sized grains to open dust gaps without perturbing the gaseous
disk significantly. However, their simulations only last 40 planetary orbits and the disk mass they adopted is far less massive
than that of a typical protoplanetary disk (and hence the Stokes number of the 1 mm dust grains becomes much larger than unity),
which makes their conclusion less plausible.

\citet[][hereafter I16]{2016PhRvL.117y1101I} presented the first conclusive study of the spatial distribution of
millimeter-sized dust particles and molecular gas in a ringed protoplanetary disk around the Herbig Ae star HD 163296 with ALMA.
The Herbig Ae star HD 163296 is a Class II young stellar object (YSO) with a stellar mass of 2.3\Msun\, and an effective
temperature of $9500\,{\rm K}$ \citep{2004A&A...416..179N}. At a distance of only $122 ^{+ 17} _{- 13} \;{\rm pc}$ from the
Earth \citep{1997A&A...324L..33V} the star has a gaseous disk extended to $\sim 550\,{\rm au}$, which makes it an ideal object
to study the gas and dust coevolution as well as planet formation in circumstellar disks at the early stage. I16 find three sets
of bright and dark rings in the dust continuum emission, similar to those observed in the HL Tau disk. Moreover, \co, \cco and
\coo J=2-1 line emission across these ringed structures are also acquired by ALMA. By comparing ALMA observations with
parameterized disk models, I16 show that the continuum rings can be explained by three dust depleted circular gaps with
depletion factors ranging from a few to 70, while CO depletion shows a contradictory result in the inner gap and consistent
results in the middle and outer gaps. Therefore, I16 conclude that the middle and outer gaps support the planet--disk
interaction scenario, however, the inner gap may require additional physical processes to explain, such as MHD instabilities and
volatile freeze-out.

Here we explore the possibility that all three gaps are formed through planet--disk interaction. In particular, the inner gap is
opened by a low-mass planet such that the gas surface density is not severely perturbed.  In the context of HD 163296 disk, we
note that the location of the inner dust gap is roughly within the magnetorotational instability 
\citep[MRI, ][]{1991ApJ...376..214B} dead zone, because the disk ionization level is predicted
to be low due to non-ideal MHD effects such as the Ohmic dissipation, Hall effects and ambipolar diffusion
\citep{2014ApJ...795...53Z,2016ApJ...821...80B}. Such a weakly turbulent disk can be parameterized with a low Shakura--Sunyaev
$\alpha$-viscosity ($\alpha \lesssim 10^{-3}$), and an embedded massive planet can trigger Rossby wave instabilities and form
visible vortices \citep{2005ApJ...624.1003L,2014ApJ...795L..39F}. Since no large-scale asymmetric feature has been detected down
to a scale as small as 25 au (I16), it is reasonable to exclude the existence of a massive planet.

In this paper, we perform \twod two fluid (gas and dust) hydrodynamic simulations to fit the observed ringed structures by I16,
which allows us to explore the general properties of the protoplanetary disk, such as the $\alpha$-viscosity. This additional
perspective is a supplement to previous efforts of turbulence measurement in the HD 163296 disk \citep[e.g.,
][]{2015ApJ...813...99F,2017ApJ...843..150F} and may help to understand the underlying physics. Our numerical model is described
in Section \ref{sec:method}. We present quantitative fitting to the ALMA observations of HD 163296 disk in terms of its
millimeter continuum and CO emission lines flux densities in Section \ref{sec:results}. We address the question whether the
planet--disk interaction model is consistent with the observation and interpret our results in terms of the strength of
turbulence in the HD 163296 disk in Section \ref{sec:discussion}.

\section{Method} 
\label{sec:method}

We follow the procedure of the HL Tau disk modeling in J16 by combining the planet--disk interaction simulations with the
\lacompass code \citep{2005ApJ...624.1003L,2009ApJ...690L..52L,2014ApJ...795L..39F} and the radiative transfer calculations with
the \radmc code \citep{Dullemond:2012vq}, except that we adopt the disk temperature proposed by I16 and we model CO emission
lines as well.

\subsection{Hydrodynamic simulation}

We adopt an initial disk dust and gas surface density profile presented in I16 which can be described by
\begin{equation}
	\Sigma(r) = \Sigma_0 \left(\frac{r}{r_0}\right)^{-\gamma}
	{\rm{exp}}\left[-\left(\frac{r}{r_{\rm{c}}}\right)^{2-\gamma}\right],
\end{equation} 
where $r_0 = 59 \,{\rm au}$. The dust surface density is characterized by $\gamma=0.1$, $r_\textrm{c} = 90$ au,
and $\Sigma_0 = 0.40 \;\textrm{g cm}^{-2}$. Similarly, the gas surface density is characterized by $\gamma= 0.8$, $r_\textrm{c}
= 165$ au, and $\Sigma_0 = 17.5 \;\textrm{g cm}^{-2}$. The disk extends from 10 to 500 au, and is simulated with a resolution of
$1024\times768$ grids along the radial and azimuthal direction.

The fixed isothermal temperature profile can be described by $T(r) = 24.1 (r/r_0)^{-0.5}$, which corresponds to 
a local sound speed that obeys 
\begin{equation}
	h(r)=\left(\frac{c_\textrm{s}}{v_\textrm{K}}\right)(r)=0.05\left(\frac{r}{r_0}\right)^{0.25}, \label{equ:cs}
\end{equation} 
where $h$ is the disk aspect ratio (disk scale height divided by radius) and $v_\textrm{K}$ is the Keplerian velocity. 
We note that the isothermal temperature profile used in hydrodynamic simulations is slightly cooler than the disk 
midplane temperature assumed in radiative transfer calculation. In Section \ref{subsec:scaleheight}, we discuss the effect 
of different scale height profiles.

In our \twod hydrodynamic simulations, the dust grains are assumed to have a uniform size of $0.15\,{\rm mm}$. This assumption
is motivated by the fact that the dust opacity at $1\,{\rm mm}$ wavelength is dominated by dust grains with sizes between 0.1
and $0.2\,{\rm mm}$ (J16).

The HD 163296 disk shows a distinctive and complex spatial distribution between the gas and the dust across different gaps.
Since viscosity has a huge impact on the evolution of a protoplanetary disk, in particular the gas depletion caused by an
embedded planet inside a gap is related to the viscosity \citep{2015ApJ...806L..15K}. Following the convention, we assume the
turbulent viscosity has the form $\nu = \alpha c_{\textrm{s}} h$ \citep{1973A&A....24..337S}, in which the Shakura-Sunyaev
$\alpha$ parameter is treated as a free parameter in this study. Both constant $\alpha$-viscosity and radial-varying
$\alpha$-viscosity disks are considered.

A planet embedded in a protoplanetary disk exerts tidal torques on the nearby disk material resulting in the angular momentum
transfer between the planet and the disk. Consequently, a planet will migrate inward or outward if it loses or gains angular
momentum from the disk. Such planet--disk tidal interaction also adjusts the disk surface density creating gaps and bumps. In
this study, however, we focus on the gap formation instead of migration. We assume that planets do not migrate for 1000 orbits
($\sim 3\times 10^5$ yr), so the gaps can be fully developed. Furthermore, we place three protoplanets at 59, 105 and 160 au,
where they are close to the centers of dust gaps. We run hydrodynamic simulations with different planet mass to search for a
model that can roughly fit the parameterized disk model proposed by I16. By doing so, we are able to narrow down the possible
sets of planet mass needed for ray-tracing calculations.

\subsection{Radiative transfer calculation of dust continuum and CO emission} 
\begin{figure}[htbp]
  \begin{center}
   \includegraphics[width= \linewidth,clip=true]{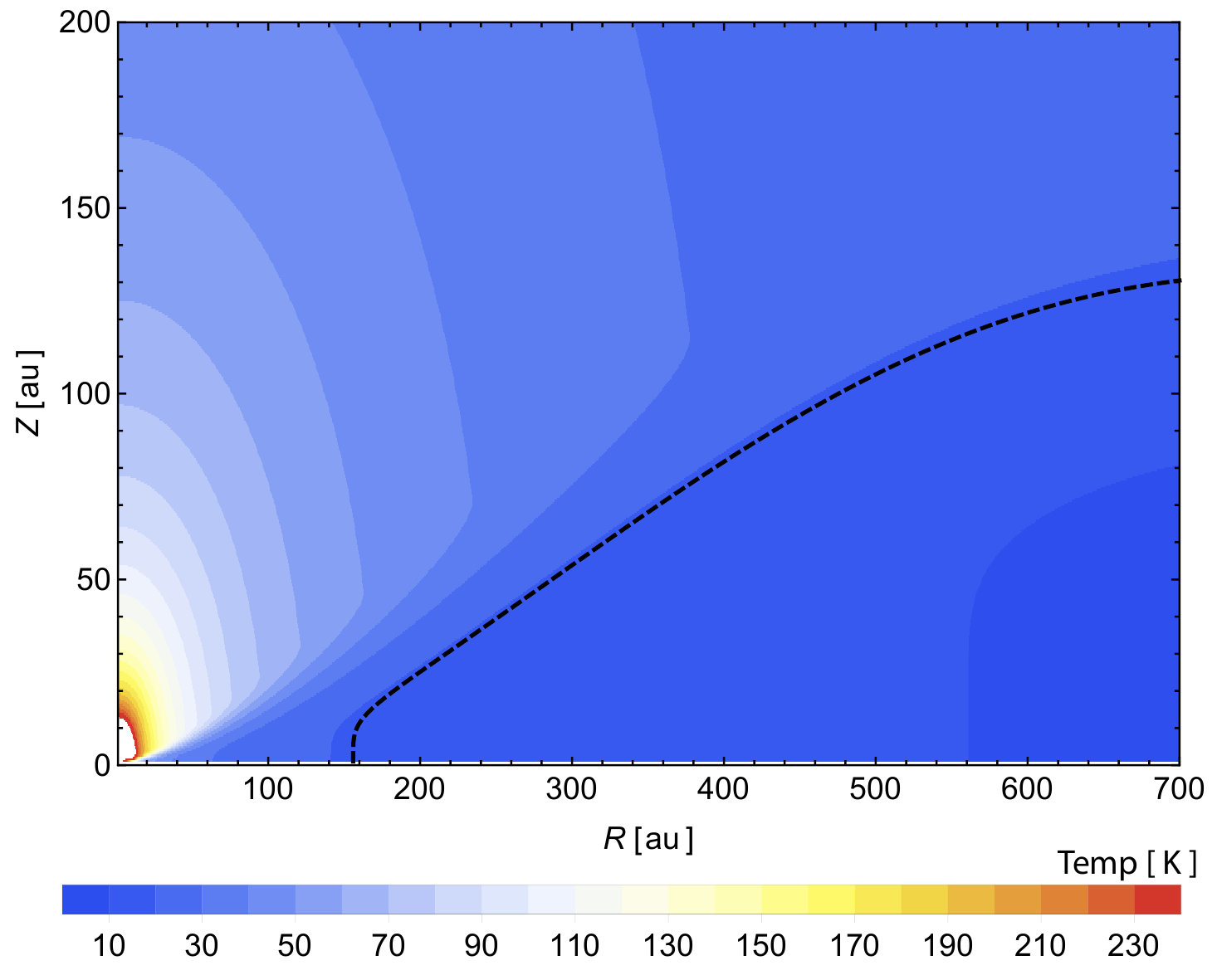}
  \caption{The contour plot of the disk temperature in the $r$--$z$ plane. The black dashed curve corresponds to the boundary
  below which CO freeze-out occurs.} \label{fig:temp} \end{center}
\end{figure}

We followed the approach described in J16 in which
hydrodynamic and radiative transfer simulations to model the dust emission of the HL Tau disk. Here we convert the dust 
surface density from a two-dimensional hydrodynamic simulation to a three-dimensional volume density assuming a dust disk scale 
height $h_\textrm{dust}(r) = 0.1\;h_\textrm{gas}(r)$. Then we generate the dust continuum emission using the radiative transfer
code \radmc \citep{Dullemond:2012vq} assuming a disk temperature profile \citep{2003A&A...399..773D} as 
\begin{equation}
 T(r,z)=\begin{cases}
		T_{\rm{a}}+\left(T_{\rm{m}}-T_{\rm{a}}\right)\left[\cos \left( \frac{\pi z}{2
		z_{\rm{q}}}\right)\right]^{2\delta},& z<z_{\rm{q}}.\\ T_{\rm{a}}(r,z),& z \geq z_{\rm{q}}.
	   \end{cases}
\end{equation} 
Here, $T_{\rm{m}}(r) = 24 \,\text{K} \,(r/100 \,\text{au}\,)^{-0.5}$ describes the disk temperature at the
mid-plane and $T_\text{a}(r,z)=45 \,\text{K}\, (\sqrt{r^2+z^2}/200 \,\text{au}\,)^{-0.6}$ depicts the disk atmospheric
temperature. Such two zones are smoothly connected with $\delta(r)=0.0034 (r-200\,\text{au}\,)+2.5$ and
$z_\text{q}=63\,\text{au}\,(r/200\,\text{au}\,)^{1.3}\exp\, [-(r/800\,\text{au}\,)^2 ]$. We plot the disk temperature profile in
the $r$--$z$ plane in Figure \ref{fig:temp}. I16 uses a similar description to interpret the continuum and line emission. 

We adopt a dust opacity at 1.3 mm of 3.95 $\textrm{cm}^2\textrm{g}^{-1}$ same as I16, which is calculated based 
on a typical dust grain composition with a grain size distribution extending up to 1 mm \citep{2009ApJ...701..260I}.

\begin{figure}[htbp]
  \begin{center}
   \includegraphics[width= 0.9\linewidth,clip=true]{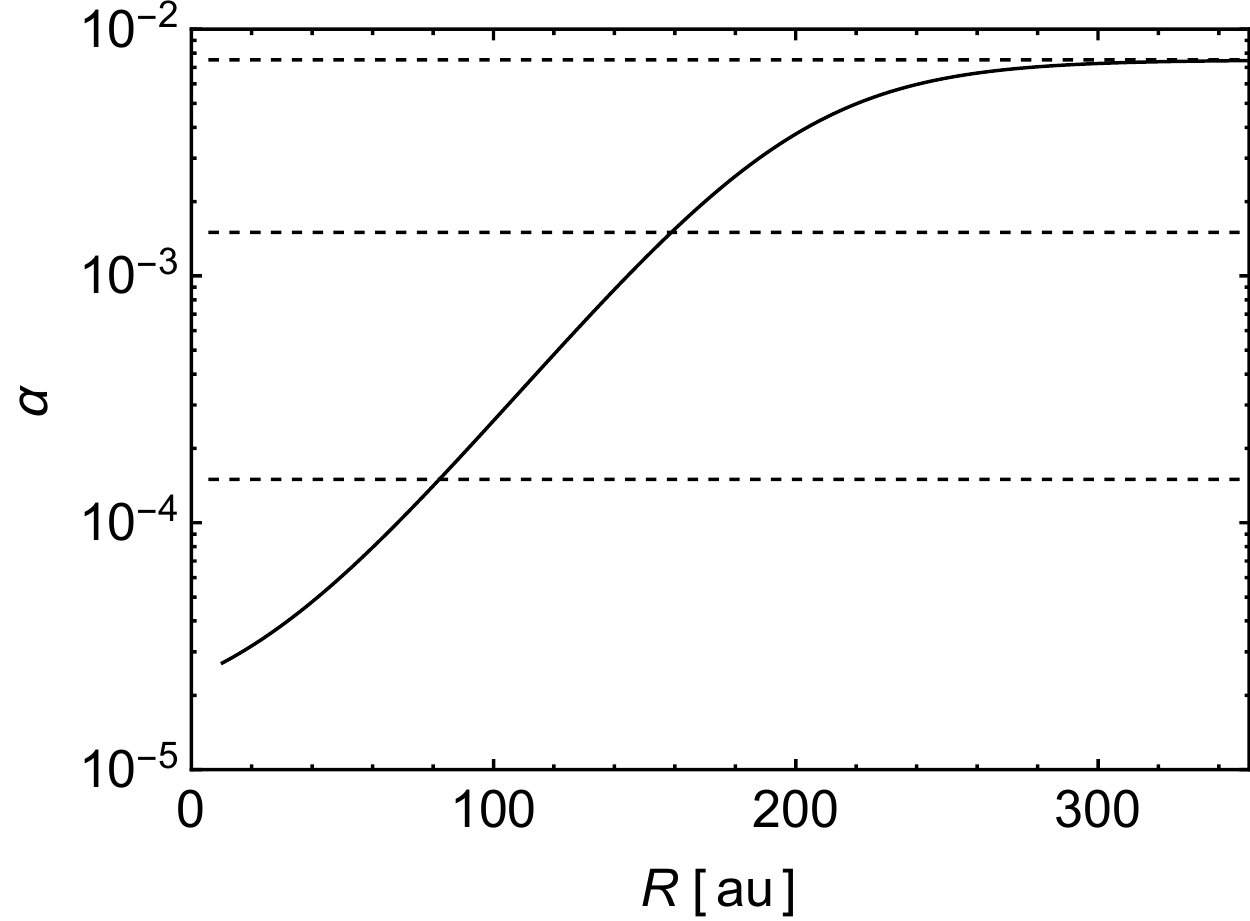}
  \caption{Effective $\alpha$-viscosity profiles used in 2D hydrodynamic simulations.
  Dashed lines represent three constant viscosity models with a high, intermediate and low $\alpha$, respectively. The solid line 
  depicts a varying $\alpha$-viscosity (Equation \ref{equ:alpha}) used in the nominal hydrodynamic model.} 
  \label{fig:alpha} 
  \end{center}
\end{figure}

We further calculate the \co, \cco and \coo emission using hydrodynamic results assuming the same temperature profile. I16
suggests that (i) the \co/ \cco and \co/ \coo density ratios are constant throughout the disk; (ii) the dust-to-gas ratio
changes with the orbital radius; (iii) and the CO / $\text{H}_2$ density ratio also varies with the orbital radius (see below).
The CO density along the vertical direction is assumed to have a gaussian distribution set by $T_\text{m}$. We reduce the CO
density by a factor of $10^8$ when the temperature is below 19 K (the black dashed line in Figure \ref{fig:temp}) to mimic the
CO freeze-out. To account for the photodissociation, the CO density is reduced by a factor of $10^8$ near the disk surface. The
gas rotational velocity is assumed to have a Keplerian speed around a central star of $2.3\, M_\odot$. The generated disk image
is inclined by $42^\circ$ and rotated by $132^\circ$ to match the observation.

The reader interested in the detailed procedure of calculating dust and CO emission using \radmc is referred to I16
supplementary material.

\section{Results} \label{sec:results}

We fix three protoplanets at 59, 105 and 160 au in all hydrodynamic simulations to investigate the planet--disk interaction and gap
properties. Normally, one can perform a grid search through the parameter space to look for good fits of the HD 163296 disk. 
Since it is a high dimensional problem, a regular grid search is not cost-effective. The ideal solution would be implementing a global 
optimization technique such as Markov chain Monte Carlo methods to search for models that fit the data. Practically, such a
method is prohibited because it requires a huge amount of computation. In this work, therefore, we improve our fitting by comparing 
our hydrodynamic simulations to the I16 fiducial model. We have performed more than one hundred simulations in total with various 
initial conditions of planet mass, disk mass and effective viscosity. Here we only show some representative cases to illustrate how we
improve our fitting and identify the nominal model.

\subsection{Constant $\alpha$-viscosity model} 
\label{subsec:const} 
\begin{figure*}[htbp]
  \begin{center}
   \includegraphics[width= 0.9\linewidth,clip=true]{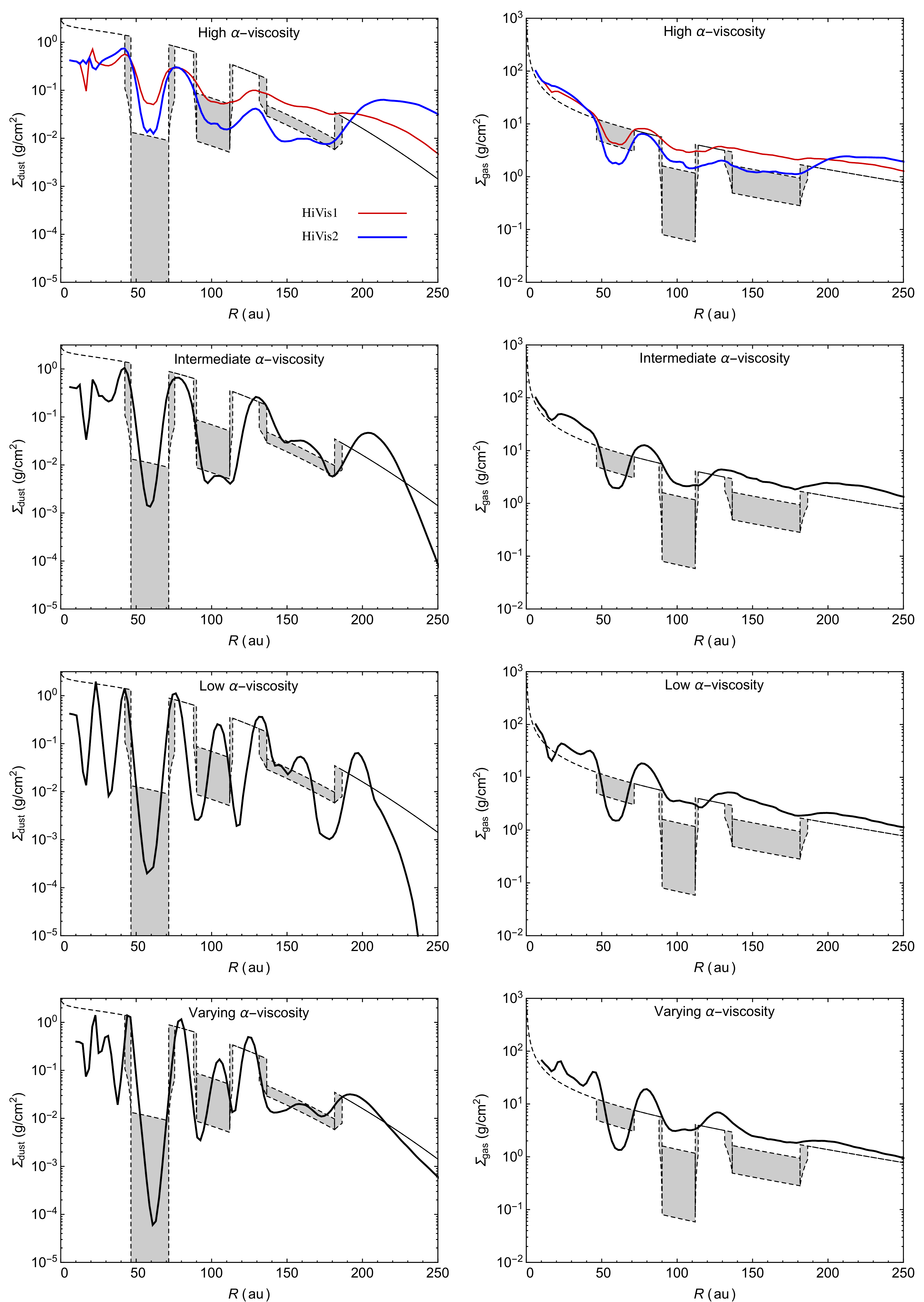}
  \caption{From top to bottom, disk models with a high viscosity ($\alpha = 7.5\times 10^{-3}$), an Intermediate viscosity
  ($\alpha = 1.5 \times 10^{-3}$), a low viscosity ($\alpha = 1.5 \times 10^{-4}$) as well as a varying viscosity. In each case,
  azimuthally averaged dust and gas surface density are shown is the left and right panels, respectively. We also plot the
  fiducial model introduced by I16 in the black dashed line. The gray shaded area indicates the uncertainty of the fiducial model.} 
  \label{fig:surfs} 
  \end{center}
\end{figure*}

First, we attempt to fit the observation with a constant
$\alpha$-viscosity disk model, which is commonly adopted in disk modeling. Here we present models with three different
alpha-viscosities $\alpha = 7.5\times 10^{-3}$, $1.5\times 10^{-3}$ and $1.5 \times 10^{-4}$, respectively (see Fig.
\ref{fig:alpha}). We plot their surface densities of dust and gas after one thousand orbits at 59 au in Fig \ref{fig:surfs}.
Parameters of disk models are listed in Table \ref{tab:sims}.

\textit{High-viscosity disk:} for a disk with the effective viscosity $\alpha = 7.5 \times 10^{-3}$, we present two sets of
simulation with different planetary mass. In the low mass planet case HiVis1, the dust gaps opened by planets are too shallow
comparing to the fiducial model suggested by I16. On the other hand, the dust surface density at large radii is relatively flat,
which is consistent with the detection of continuum emission of the HD 163296 disk at a large angular distance. If we increase
the planet mass (HiVis2 case), we do get deeper dust gaps. However, massive planets severely hinder dust drift forming a giant
bump of the dust surface density outside the furthest gap. Besides, the second and third gaps are not well separated in dust and
gas surface density in both cases.

\textit{Intermediate-viscosity disk:} if we lower the effective viscosity to $\alpha = 1.5 \times 10^{-3}$ (InVis case), we can
get a satisfactory fit of the middle and the outer dust gap, while the inner dust gap is slightly narrower than the fiducial model.
The major drawback of this model is the depletion of dust beyond 200 au (see the second row in Figure \ref{fig:surfs}).

\textit{Low-viscosity disk:} With a even lower viscosity $\alpha = 1.5 \times 10^{-4}$ (LoVis case), we can fit the inner gap
better than the intermediate-viscosity case. However, the fitting of the middle and the outer gap becomes worse. Again, the fast 
drop-off of dust surface density is not consistent with the observation (see the third row of Figure \ref{fig:surfs}).

One interesting trend is that lower $\alpha$ viscosity leads to a more compact dust disk at the end of our simulation. 
The reason behind this phenomenon is the disk viscous evolution. Our initial gas surface density has a exponential cut-off 
$r_\textrm{c}$ at 165 au, which is just inside the outer gap. As the disk evolves, angular momentum is transported outward, so 
the disk spreads out and the characteristic scale $r_\textrm{c}$ increases. When the viscosity is high, the gas disk spreads out 
so fast, such that the dust inward drift could be balanced. While in the low-viscosity disk, the angular momentum transport is 
inefficient, dust grains quickly get depleted in the outer disk. 

To conclude, we find it is difficult to reproduce all the features suggested by the observation with a constant
$\alpha$-viscosity disk.

\subsection{Varying $\alpha$-viscosity model} 
\label{subsec:nominal} 

\begin{deluxetable*}{llllccclll}
  \tabletypesize{\footnotesize} 
  \tablecolumns{8} 
  \tablewidth{0pt} 
  \tablecaption{Disk Parameters of Hydrodynamic Simulations\label{tab:sims}}
  \tablehead{\colhead{Model}            &
	   \colhead{$M_\text{p1} \,(M_\text{J})$}       &
       \colhead{$M_\text{p2} \,(M_\text{J})$}   & \colhead{$M_\text{p3} \,(M_\text{J})$}   & \colhead{$\alpha$} &
       \colhead{$\alpha_\text{in}$} & \colhead{$\alpha_\text{out}$} & \colhead{$\sigma$} & \colhead{$R$ (au)} & \colhead{$h(r)$} }
   \startdata
   \textit{Constant $\alpha$:}\\
   HiVis1 & 0.69 & 0.69 & 0.69 & $7.5 \times 10^{-3}$ & \NA & \NA & \NA & \NA & $0.05\times(r/r_0)^{0.25}$\\ 
   HiVis2 & 1.84 & 1.84 & 2.30  & \ditto & \NA & \NA & \NA & \NA & \ditto\\ 
   InVis & 0.92 & 0.92 & 0.69 & $1.5 \times 10^{-3}$  & \NA & \NA & \NA & \NA & \ditto\\ 
   LoVis & 0.46 & 0.46 & 0.35 & $1.5 \times 10^{-4}$  & \NA & \NA & \NA & \NA & \ditto\\ 		
   \textit{Varying $\alpha$:}\\
   Nominal & 0.46 & 0.46 & 0.58 & $\alpha(r)$ & $10^{-5}$  & $7.5\times 10^{-3}$ & 1 & 155 & $0.05\times(r/r_0)^{0.25}$\\
   VaVis1 & 0.69 & 0.46 & 0.69 & $\alpha(r)$ & $5\times 10^{-5}$ & \ditto & 1 & 155 & \ditto\\ 
   VaVis2 & 0.23 & 0.40 & 0.52 & $\alpha(r)$  & $10^{-6}$ & \ditto & 1 & 155 & \ditto\\ 
   ShallowT & 0.46 & 0.46 & 0.80 & $\alpha(r)$ & $10^{-5}$  & \ditto & 1 & 155 & $0.05\times(r/r_0)^{0.35}$\\
   LargeH$_0$ & 0.69 & 0.80 & 1.04 & $\alpha(r)$ & $10^{-5}$ & \ditto & 1 & 155 & $0.057\times(r/r_0)^{0.25}$
   \enddata 
   \tablecomments{See Equation \ref{equ:alpha} for the definition of $\alpha(r)$.}
\end{deluxetable*}

From the observation we know that the inner dust gap is much more prominent than the inner gas gap (if there is any depletion in
the gas disk). And a planet with mass as low as $0.46 \; M_\text{J}$ embedded in a low-viscosity disk can effectively stop 0.15
mm-sized dust particles from drifting inward in the inner part of the disk and open a sufficiently deep dust gap. On the other
hand, such a low mass planet does not deplete the gas component as much as it does to the dust. The different dust-to-gas ratio
across the three gaps found by I16 suggests that varying planet mass alone cannot account for the whole picture.

Inspired by the fact that each constant $\alpha$-viscosity disk model presented in Section \ref{subsec:const} can explain part
of the disk but fails in reproducing all the features, we come up with a radius-depended effective viscosity 
\begin{equation}
	\alpha (r) = \alpha_\textrm{in} \left[1-\frac{1-\alpha_\textrm{out}/\alpha_\textrm{in}}{2}
	\left(1-\tanh\frac{r-R}{\sigma r_0}\right)\right], \label{equ:alpha}
\end{equation} 
where $\alpha_\textrm{in}$ and $\alpha_\textrm{out}$ are dimensionless numbers which denote a low
viscosity ($\alpha \lesssim 10^{-4}$) in the inner disk and a large viscosity ($\alpha \sim 0.01$) in the outer disk,
respectively. Parameter $R$ sets the mid-point of the transition and $\sigma$ controls the slope. The viscosity profile
effectively mimics a MRI dead zone near the mid-plane of the inner disk \citep{2017ApJ...835..118M}, while approaches a larger
quantity where MRI can operate. A similar $\alpha$ prescription that includes a dead zone is also proposed by
\citet{2016A&A...596A..81P}.

In the best-fit nominal model, we choose  $\alpha_\textrm{in} = 10^{-5}$ and $\alpha_\textrm{out} = 7.5 \times 10^{-3}$, which
are the asymptotic effective $\alpha$-viscosity at the infinity and at the origin, respectively. Other parameters are $\sigma =
1$, $r_0 = 59$ au and $R = 155$ au. The effective $\alpha$-viscosity as a function of radius is plotted in Figure
\ref{fig:alpha}. \citet{2016ApJ...821...80B} assumed that disk evolution is wind-driven and proposed an effective viscosity
model that is qualitatively similar but quantitatively different from ours, which has a larger $\alpha_\textrm{in}$ and smaller
$\alpha_\textrm{out}$ with a less steep slope compared to our model.

\begin{figure*}[htbp]
  \begin{center}
   \includegraphics[width= 0.95\linewidth,clip=true]{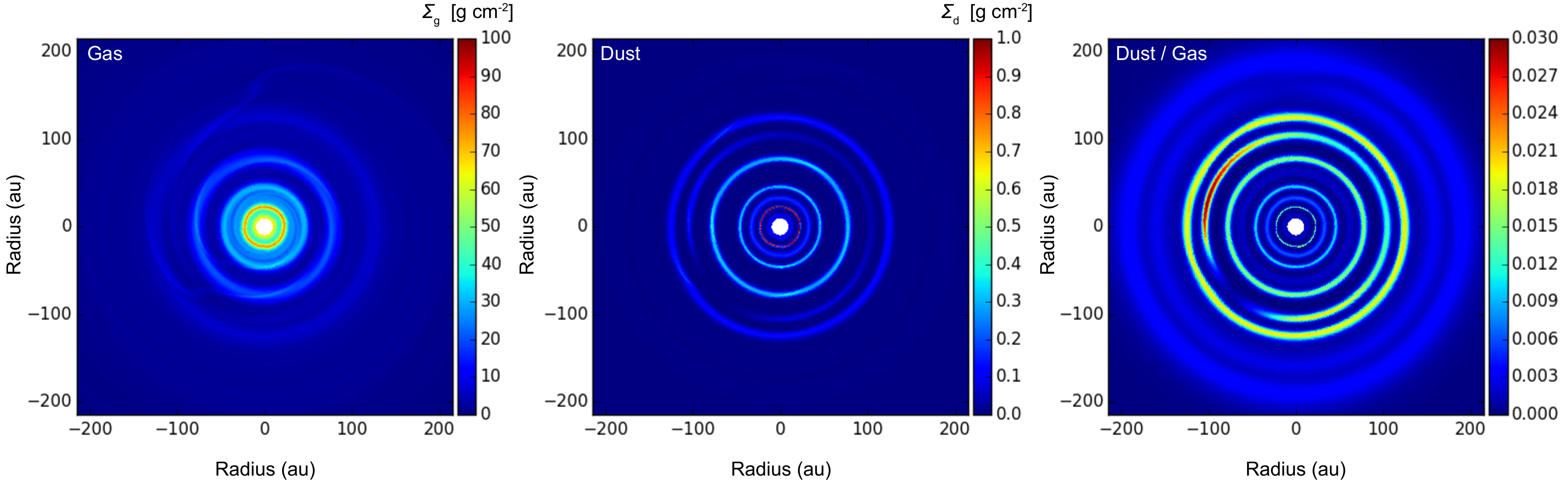}
  \caption{Gas surface density (left) and \textbf{dust} surface density (middle) and dust-to-gas ratio (right) of the nominal 
  model after 1,000 orbits at 59 au.} 
  \label{fig:hydro} 
  \end{center}
\end{figure*}

We fix three half-Jovian-mass planets (0.46, 0.46 and 0.58 Jupiter masses) at 59, 105 and 160 au in the simulation. Figure
\ref{fig:hydro} shows the surface density of gas and 0.15 mm-sized dust obtained after one thousand turns of the inner planet.
Such a choice allows gaps created by the planet--disk tidal interaction to be fully developed. To compare the nominal model with
the parametric model in I16, we plot the one-dimensional surface density of gas and dust of the two models in Figure
\ref{fig:surfs}. Overall, the planet--disk interaction model can create dust gaps comparable to those in parametric models in
terms of the depth and width, while the inner gas gap of the nominal model is deeper than that in the parametric models.  Note
that the hydrodynamic model is more realistic than the parametric models on characterizing gaps. In particular, gaps created by
planets have smooth transitions rather than sharp edges in parametric models.

\begin{figure*}[htbp]
  \begin{center}
   \includegraphics[width= \linewidth,clip=true]{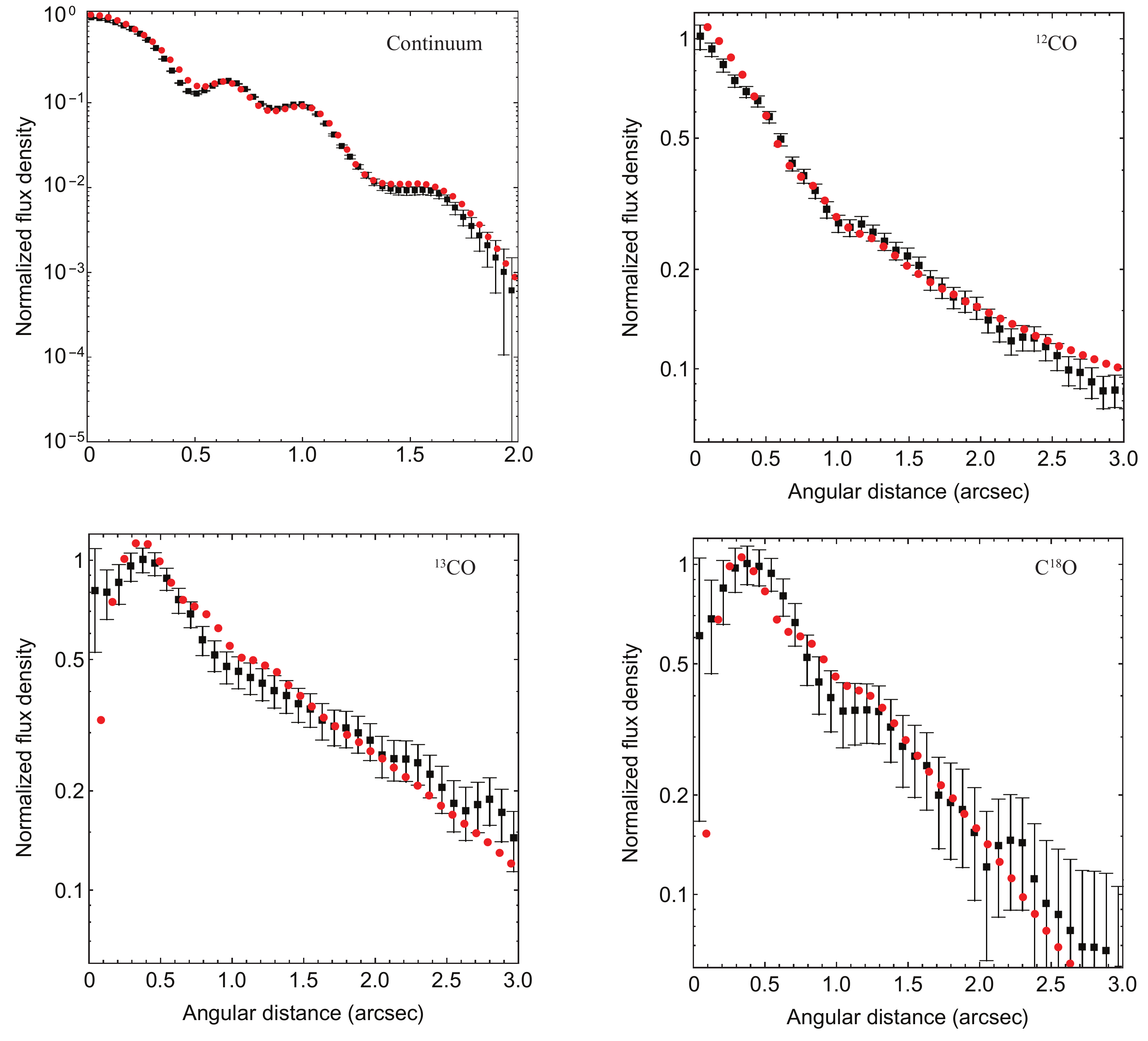}
  \caption{Comparison between the observation (black dots) and the hydrodynamic nominal model (red dots) for the dust continuum,
  \co, \cco and \coo emission, respectively.} 
  \label{fig:radmc_lines} 
  \end{center} 
\end{figure*}

\begin{figure*}[htbp]
  \begin{center}
   \includegraphics[width= \linewidth,clip=true]{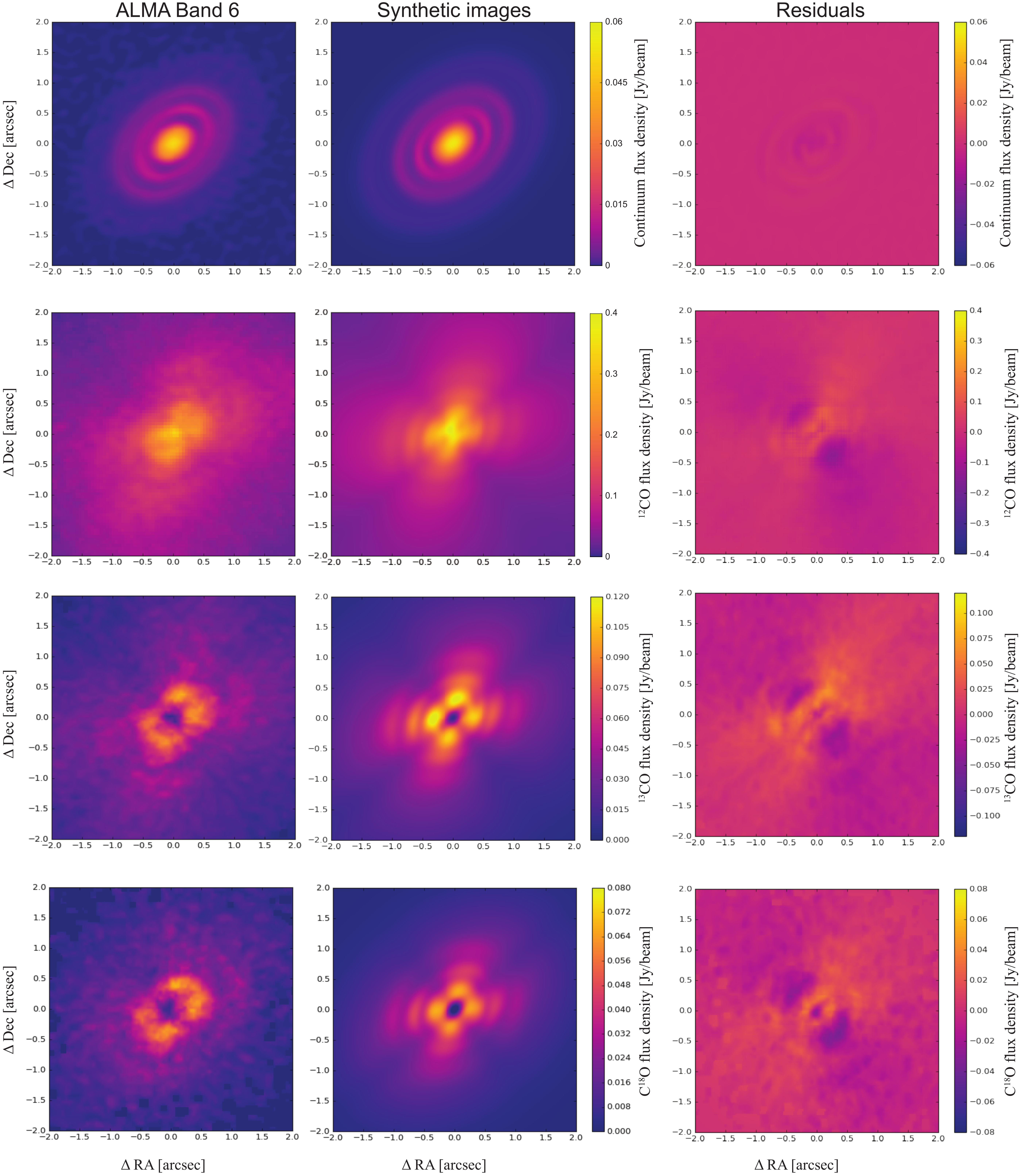}
  \caption{Comparison between ALMA's images of the HD 163296 disk emission in 1.3 mm (left column) and the synthetic emission of
  the nominal model convolved with the synthesized beam of the observation (central column). The right column shows the residual
  of each map, i.e., the difference between our nominal model and the observation. The top row shows the continuum flux density,
  while the lower three rows are the continuum-subtracted \co, \cco and \coo flux density, respectively.}
  \label{fig:radmc_images} 
  \end{center} 
\end{figure*}

Furthermore, we run the ray-tracing of our nominal model to compared with the observation. Since the disk is axisymmetric (see
Figure \ref{fig:hydro}), we plot the azimuthally averaged flux density normalized to the peak intensities of dust continuum at
1.3 mm and CO J=2-1 emission lines in red dots in Figure \ref{fig:radmc_lines}. We also compare the synthetic dust continuum and
continuum-subtracted CO J=2-1 emission of the nominal model to the ALMA images of HD 163296 disk in figure
\ref{fig:radmc_images}.

Overall, our nominal model reproduces the observation. Three dust gaps in our nominal model create ringed structures that
matches the observation (see the upper-left panel of Figure \ref{fig:radmc_lines} as well as Figure \ref{fig:radmc_images}). On
the other hand, the other three maps of CO emission lines in Figure \ref{fig:radmc_lines} do not clearly show any deficit at the
location of the inner dust gap (at about 0.5 arcsec) and most of the data points produced by the nominal model are within the
error bars.

Albeit the inner gas gap of the nominal model is much deeper than that of the I16 disk model (see the right panel of Figure
\ref{fig:surf}), ringed structures are not manifested in the synthetic image of CO emission. As mentioned above, the inner gas
gap of the nominal model has less steep edges and a narrower bottom compared to the parametric model. Furthermore, the
"shoulders" of the inner gas gap have a higher surface density than that in the parametric model, which offsets the gas
depletion within the gap. Besides, the disk is substantially inclined and such a viewing angle can mitigate the effects of a
deep narrow gap. As a result, the ringed structures are barely discernible on maps of CO emission, not to mention that the
convolution with a Gaussian beam of observations will average out the fluctuations of CO emission if there are any.

\subsection{Constraints on $\alpha$-viscosity and the mass of the inner planet} 
\label{subsec:alpha} 
\begin{figure*}[htbp] 
 \begin{center}
 \includegraphics[width= \linewidth,clip=true]{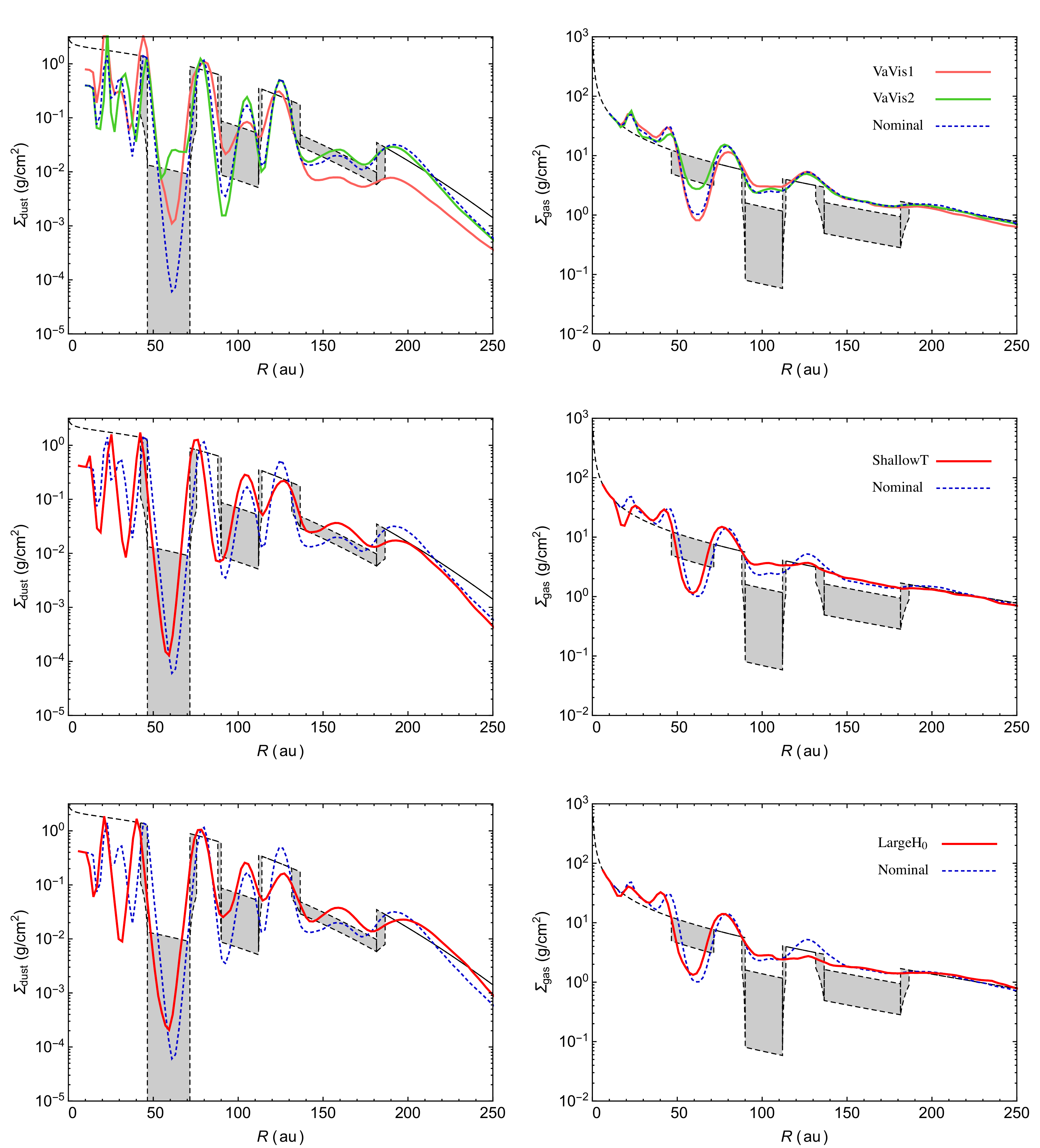}
\caption{From top to bottom, azimuthally averaged dust surface density (left) and gas surface density (right) of disks 
with slightly different varying $\alpha$ profiles (top) and modified scale height profiles (middle and bottom). Parameters of 
all models are listed in Table \ref{tab:sims}. The nominal model is shown in blue dashed line for reference. We also plot the 
I16 fiducial model in black dashed line as well as its uncertainty represented with the gray shaded area.} 
\label{fig:surf} 
\end{center} 
\end{figure*}

Through testing a series of constant viscosity disk models, we demonstrate that none of those models works. Based on a 
varying $\alpha$ viscosity, we are able to construct disk models that are similar to the fiducial model. However, fitting 
three dust and gas gaps simultaneously requires additional fine-tuning. The nominal model described in the previous subsection 
is the best fit among our simulations. To show how sensitive our results depend on the $\alpha$-viscosity and planet mass (in  
particular the inner one), we plot the dust and gas surface density of another two hydrodynamic simulations in colored solid lines 
in the top row of Figure \ref{fig:surf}.

In model VaVis1 we use the same disk as in the nominal model except that we choose $\alpha_\textrm{in} = 5 \times 10^{-5}$. Because the
$\alpha$-viscosity is slightly larger than that of the nominal model in the inner the disk, even a more massive planet 
(0.69 $M_\textrm{J}$) does not open a gap as deep as the gap opened by a less massive planet (0.46 $M_\textrm{J}$) in the nominal
model. However, both of them are considered to be good fits compared to the I16 parametric disk model. The fact that there is a
degeneracy between the $\alpha$-viscosity and the inner planet mass is because we do not know exactly the dust depletion in the
first gap, as the observation only constrains the upper limit.

However, the fitting is more sensitively to the mass of the inner planet than the $\alpha$-viscosity as long as $\alpha_\textrm{in}$ 
being very low. For example, in model VaVis2 we adopt $\alpha_\textrm{in} = 1 \times 10^{-6}$ and a 0.23 $M_\textrm{J}$ planet. 
Although we lower the $\alpha$-viscosity, such a low mass planet still cannot open a wide and deep gap in the dust disk. Besides, 
we also decrease the mass of the other two planets. Consequently, the fitting of the other two gaps becomes worse.

Hence, our parameter space study suggests that the inner gap can be created by a sub-Jovian mass planet with a mass in the
range of 0.46 to 0.69 $M_\textrm{J}$ in a low-viscosity environment ($\alpha \lesssim 10^{-4}$). The mass of the middle and the
outer planet is well constrained given the $\alpha$-viscosity. Based on our results, we exclude the possibility of $\alpha$
being greater than $10^{-2}$ in the outer disk ($r > 300$ au).

\subsection{Constraints on the disk scale height}
\label{subsec:scaleheight}
\begin{figure*}[htbp] 
 \begin{center}
 \includegraphics[width= \linewidth,clip=true]{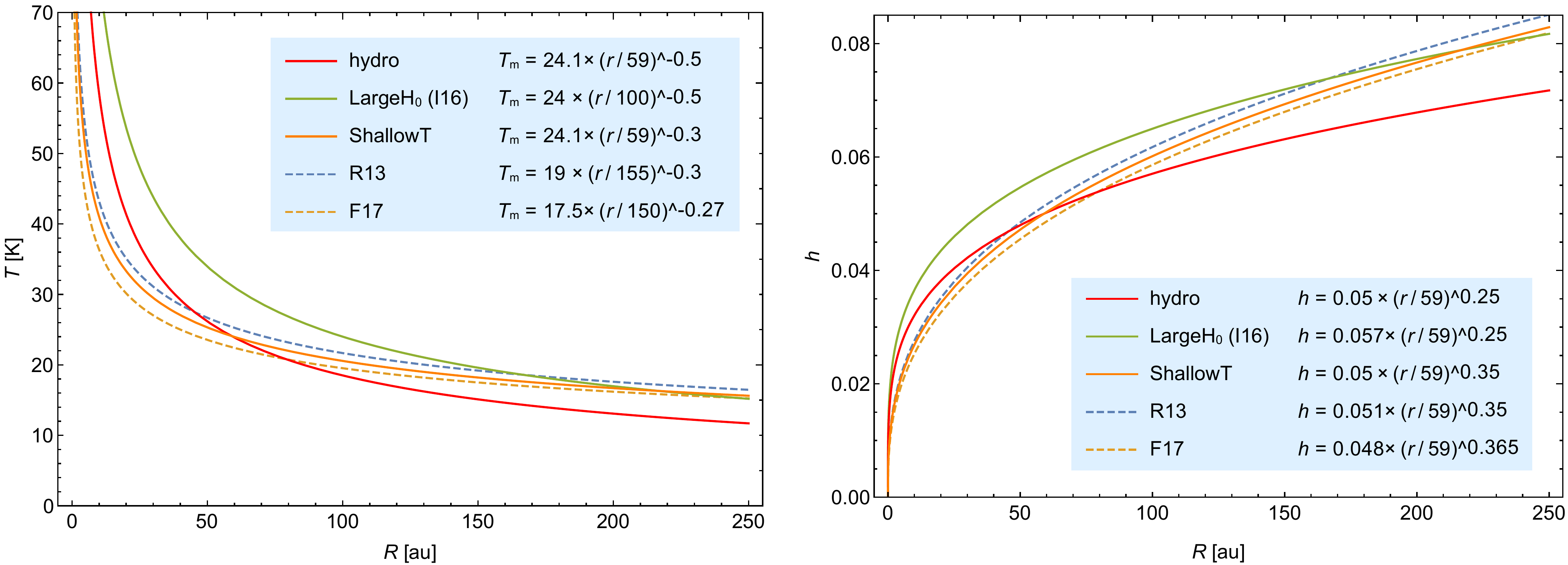}
\caption{Comparison of midplane temperature profiles (left) and disk aspect ratios (right) of various disk models. 
The hydro model stands for the isothermal temperature profile assumed in our hydrodynamic simulations, except two additional 
hydrodynamic simulations ShallowT and LargeH$_0$ presented in Section \ref{subsec:scaleheight}. We also plot midplane temperature 
profiles and disk aspect ratios from \citet{2013ApJ...774...16R} and \citet{2017ApJ...843..150F} in dashed lines labeled as R13 and F17, respectively.}
\label{fig:scaleheight} 
\end{center} 
\end{figure*}

Hydrodynamic models presented in previous sections are all assumed a disk aspect ratio $h(r) = 0.05\times(r/r_0)^{0.25}$, 
corresponding to a midplane temperature $T(r) = 24.1 \times (r/r_0)^{-0.5}$. The power index is same as I16, while the normalization 
is slightly different. Other studies \citep{2013ApJ...774...16R, 2017ApJ...843..150F} also suggest temperature profiles with power-law 
indices shallower than $-0.5$. We plot various midplane temperature profiles and corresponding disk aspect ratios in Figure 
\ref{fig:scaleheight}. Here, we explore the impact on the fitting due to different temperature profiles.

First, we run hydrodynamic simulations with disk aspect ratio $h(r) = 0.05 \times (r/r_0)^{0.35}$ corresponding to a disk 
midplane temperature $T(r) = 24.1 \times (r/r_0)^{-0.3}$. The result of the hydrodynamic simulation labeled as ShallowT is plotted 
in the middle row of Figure \ref{fig:surf}. We find the new scale height profile has little impact on the inner gap since the scale 
height $h_0$ keeps the same. At larger radii, the disk scale height becomes larger than the nominal model (see the right panel of 
Figure \ref{fig:scaleheight}). We therefore increase the planet mass from 0.58 $\textrm{M}_\textrm{J}$ to 0.80 
$\textrm{M}_\textrm{J}$ to fit the outer gap, but the gas gap is still much shallower than the nominal model.

Then, we investigate an overall larger scale height profile by increasing $h_0$ only, i.e. $h(r) = 0.057 \times (r/r_0)^{0.25}$, 
corresponding to the I16 midplane temperature, which is also the assumed midplane temperature in our RADMC-3D calculations. Since the 
new scale height is alway greater than the nominal model, we increase the mass of all three planets accordingly (Table \ref{tab:sims}). 
The model LargeH$_0$ is plotted in the bottom row of Figure \ref{fig:surf}. The inner gap matches reasonably well, while the outer gap, particularly the gas gap, is less prominent than that in the nominal model.

Disk modeling sensitively depends on the temperature, which is usually poorly constrained. Through the two additional simulations 
with different scale height profiles, we show the planet--disk interaction model still works within the uncertainty of temperature, 
but extra fine-tuning of planet mass might be needed. On the other hand, assumed temperature profiles also affect the estimation of disk turbulence. And we shall discuss its impact in Section \ref{subsec:nonidealmhd}.

\section{Discussion} 
\label{sec:discussion} 
In this study, we have found that the dust and gas features observed in the HD
163296 disk can be explained in the framework of planet--disk interaction models. However, this requires assuming that the
effective viscosity $\alpha$ increases across the disk radial extent by more than two orders of magnitude, varying from about
$3\times10^{-5}$ at 10 au to about $10^{-2}$ at 300 au. We first discuss the relationship between the $\alpha$-viscosity and
mass accretion rate, and then we try to explain what underlying physics is responsible for such a steep increase in the
$\alpha$-viscosity.

\subsection{Viscosity and mass accretion rate} In a steady thin accretion disk, the mass accretion rate at a radial distance $r$
can be expressed as 
\begin{equation}
  \label{equ:acc} \dot{M} = -2\pi r \Sigma v_r,
\end{equation} 
where $\Sigma(r, t)$ is the disk surface density and $v_r$ is the radial velocity of the gas. Now we need to work
out $v_r$. We write down the continuity equation (mass conservation equation) in the 2D form \citep[see, e.g. Chapter 5 of
][]{Frank:2002tf} 
\begin{equation}
  \label{equ:mass} \frac{\partial\Sigma}{\partial t} + \frac{1}{r}\frac{\partial}{\partial r} \left(\Sigma r v_r \right) = 0,
\end{equation} 
and the angular momentum conservation equation 
\begin{equation}
  \label{equ:ang} \frac{\partial}{\partial t}\left(\Sigma v_\phi r \right) + \frac{1}{r}\frac{\partial}{\partial r}\left(\Sigma
  v_r v_\phi r^2\right) = \frac{1}{r}\frac{\partial}{\partial r}\left(\Sigma \nu r^3 \frac{\partial\Omega}{\partial r} \right),
\end{equation} 
where $v_\phi$ is the tangential velocity of the gas, $\nu$ is the kinematic viscosity, and $\Omega = v_\phi / r
$ is the angular velocity.

Combining Equations \ref{equ:mass} and \ref{equ:ang}, we get 
\begin{equation}
  \Sigma v_r \frac{\partial}{\partial r}\left(\Omega r^2\right) = \frac{1}{r}\frac{\partial}{\partial r}\left(\Sigma \nu r^3
  \frac{\partial\Omega}{\partial r} \right).
\end{equation} 
Since the real gas velocity is very close to the Keplerian velocity, we can replace $\Omega$ with a Keplerian
angular velocity $\Omega_\textrm{K}$ , then we can get 
\begin{equation}
  v_r = -\frac{3}{\Sigma \sqrt{r}} \frac{\partial}{\partial r} \left(\Sigma \nu \sqrt{r} \right).
\end{equation}

In the inner part of the disk where the exponential term is not significant, $\Sigma$ can be approximated to a power-law
function $\Sigma \sim r^{-\gamma}$. We have assumed the sound speed has the form $c_\textrm{s} \sim r^{-q}$, so $\nu = \alpha
c_\textrm{s} h = \alpha c^2_\textrm{s}/\Omega_\textrm{K} \sim r^{-2q+3/2}$. Thus, we get \begin{equation}
  \label{equ:vr} v_r = -3(2-2q-\gamma)\frac{\nu}{r}.
\end{equation}

We can further assume that $\Sigma$ does not depend on $t$. The continuity Equation \ref{equ:mass} becomes 
\begin{equation}
  \frac{\partial}{\partial r} \left(\Sigma r v_r\right) = 0.
\end{equation} 
Then we get the relation between the power indices 
\begin{equation}
  \gamma = -2q + 3/2.
\end{equation} 
Hence the Equation \ref{equ:vr} becomes \begin{equation}
  \label{equ:vr2} v_r = \frac{3}{2} \frac{\nu}{r}.
\end{equation} 
Finally, we plug Equation \ref{equ:vr2} in for the mass accretion rate Equation \ref{equ:acc}, and get
\begin{equation}
  \label{equ:acc2} \dot{M} = 3\pi\alpha\frac{c^2_\textrm{s}}{\Omega_\textrm{K}}\Sigma.
\end{equation}

Using equation \ref{equ:acc2}, we estimate a mass accretion rate 
$\dot{M} \simeq  3 \times 10^{-11} \;\textrm{M}_\odot\;\textrm{yr}^{-1}$ at 0.1 au by extrapolating the parameters adopted in 
our model toward the disk inner edge.

The mass accretion rate of Herbig Ae/Be star HD 163296 was proposed around $10^{-7} \;\textrm{M}_\odot \;\textrm{yr}^{-1}$ \citep[see,
e.g.][]{2006A&A...459..837G, 2011A&A...535A..99M}. Before comparing the observation with the theory, we need to emphasis
the fact that Herbig Ae/Be stars have a large optical spectroscopic variability, e.g. the equivalent width of $\textrm{H}\alpha$ line 
can change by a factor of four \citep{2011A&A...529A..34M}. Such a large variability certainly confounds any comparisons between 
viscosity, surface density and the accretion rate.

Nonetheless, the order-of-magnitude difference in the mass accretion rate can be understood from two angles. First, the mass 
accretion rate of Herbig Ae/Be stars is estimated using the magnetospheric accretion model \citep{Uchida:1985vu} that has 
been applied to T Tauri stars. However, \citet{Alecian:2012ke} carried out searching for magnetic field among a group of 
Herbig Ae/Be stars. And they were able to find only 5 out of 70 Herbig Ae/Be stars to be magnetic. It is unclear how reliable 
the mass accretion rate of Herbig Ae/Be stars estimated using the magnetospheric accretion model is. Furthermore, 
\citet{2018ApJ...852....5R} found that emission line profiles are similar between magnetic and non-magnetic Herbig Ae/Be stars, 
suggesting that magnetospheric accretion is not the source of the line profile shape.

Second, our effective $\alpha$-viscosity profile is constructed based on the dust-to-gas ratio across the three gaps. The
extrapolation of $\alpha$ to the inner edge of the disk may not hold, because the very inner part of the disk, where the disk
midplane could be very hot ($>10^3$ K), may be subject to MRI and become very turbulent. In that case, the $\alpha$ value could
become much larger, and the mass accretion rate could be orders of magnitude larger than our estimation. The boundary between
the putative inner MRI-active zone and the MRI-dead zone inferred from our disk modeling is not clear. Future observations with
a higher angular resolution (see next subsection for details) may help to settle this issue.

\subsection{Non-ideal MHD effects and disk turbulence}
\label{subsec:nonidealmhd}
The radial increase of $\alpha$ is consistent with the conventional 
understanding of MRI in protoplanetary disks. In particular, MRI would not 
operate in the most dense and neutral inner disk regions. It is generally assumed that protoplanetary disks are mostly neutral
at column density larger than 10 $\textrm{g}/\textrm{cm}^2$ (though this number depends on the specifics of disk ionization
processes). Interestingly, in the case of HD 163296, this surface density level is achieved right at the inner edge of the
innermost gap for a large range of models (see right panel of Figure 3.). Numerical simulations have shown that MRI
could be quenched by non-ideal MHD effects, such as Ohmic dissipation \citep{1996ApJ...457..798J}, Hall effect 
\citep{1999MNRAS.307..849W}, and ambipolar diffusion \citep{1994ApJ...421..163B}.

According to the conventional picture, Ohmic dissipation dominates the region near the mid-plane of the innermost disk (say 1-5
au), where the density is high and the magnetic field is weak. However, due to the large optical depth of the emission arising
from the innermost part of a disk, ALMA is unlikely to place observational constraints on the turbulence in these regions 
given its large angular resolution. One possible way is moving to shorter wavelengths. \citet{2004ApJ...603..213C} find evidence 
for transonic turbulence within 0.3 au of the disk around the young star SVS 13 based on the broadening of the water lines at 
infrared wavelengths, while \citet{2017ApJ...847....6N} suggest that mechanical heating from turbulence in the inner disk may be 
observable at UV wavelengths. On the other hand, future radio interferometers such as the Next Generation Very Large Array
\citep{Carilli:2015vd,Isella:2015wa} might allow us to study the planet--disk interaction on a scale as small as 1 au 
\citep{2018ApJ...853..110R}, and, through an analysis similar to what presented in this paper, information about the 
effect of Ohmic dissipation might be inferred.

The ambipolar diffusion, on the other hand, operates under the condition that the density is low while the magnetic field is
strong. Typically, the ambipolar diffusion plays a crucial role high above the disk mid-plane \citep{2008MNRAS.388.1223S}. Our
effective viscosity $\alpha$ profile is based on \twod hydrodynamic simulations and an isothermal equation of state. 
Such a treatment simplifies the vertical structure of a disk with a disk scale height profile. Our effective 
$\alpha$-viscosity profile should mainly represent constraints near the disk mid-plane. Thus, it is not clear whether 
the decreasing $\alpha$ profile with radius is related to the variation of the effectiveness of the ambipolar diffusion. Alghouth it 
is possible to model the disk ionization driven by far-ultraviolet photons at different locations of a disk in a semi-analytical way 
\citep[see, e.g.][]{2011ApJ...735....8P}. Future three-dimensional MHD simulations are crucial to identify the ambipolar diffusion 
regime in the disk.

The Hall effect might be responsible for the low-$\alpha$ region near the inner gap found in our effective viscosity profile,
since it dominates the other two non-ideal MHD effects under a wide range of conditions \citep[see, e.g.,
][]{2011ARA&A..49..195A,2014prpl.conf..411T}. The Hall effect is estimated to produce an effective viscosity as low as 
$\alpha \sim 10^{-5}$ \citep{2014prpl.conf..411T}. Recent three-dimensional shearing box simulations show that the Hall effect could 
reduce the MRI turbulence to $\alpha \sim 10^{-3}$ \citep{2015ApJ...798...84B,2015MNRAS.454.1117S}.

Our nominal model also suggests that the effective viscosity climbs steadily until reaching a maximum at the location between 200
and 300 au. This is because non-ideal MHD effects become weaker at larger radii since the disk density decreases and
nonthermal ionization level increases. Besides, due the nature of a flaring disk, the disk surface gets more directly illuminated 
by the central star at larger radii, which results in a deeper MRI-active layer.

Independent constraints on the gas turbulence in the HD163296 disk were obtained by several studies through analysis of 
multiple molecular lines. \citet{2016MNRAS.461..385B} use CO observations to constrain the $\alpha$ of the order of $10^{-3}$ within 
90 au of the disk, which is in good agreement with our result. \citet{2015ApJ...813...99F,2017ApJ...843..150F} 
find that CO isotopes and DCO+ emissions 
\footnote{Different emission lines have different optical depth. To make a fair comparison with our $\alpha$-viscosity derived 
from the 2D simulations, one might choose emission lines that is emitted near the mid-plane. For instance, DCO+(3-2) and \coo(2-1) 
are better tracers than CO(2-1) and CO(3-2), because DCO+ and \coo are less abundant and emissions of DCO+(3-2) and \coo(2-1) come 
from regions closer to the mid-plane.}
are consistent with gas turbulence velocity less than a few percent of the sound speed, corresponding to $\alpha \lesssim 10^{-3}$. 
In particular, \citet{2017ApJ...843..150F} suggest weak turbulence ($\alpha \lesssim 3\times10^{-3}$) in the outer disk 
($r$ > 260 au), which seems to be contrary to our conclusion based on the planet--disk interaction model. 

One might explore other possibilities such as condensation fronts of different molecules \citep{2015ApJ...813..128Q} to explain the 
ringed structures in the HD 163296 disk. Here, however, we try to understand the discrepancy of $\alpha$-viscosity in the outer disk 
between our result and that in \citet{2017ApJ...843..150F}. One big assumption in \citet{2015ApJ...813...99F,2017ApJ...843..150F} 
is that the turbulent velocity dispersion is a constant fraction of the local sound speed throughout the disk, which may not be the case 
if MRI is operational. As a result, they are essentially deriving an intensity-weighted average value of $\alpha$ over the entire disk. 
We suspect that a different parameterization of $\alpha$ as a function of radius can yield a similar result following their procedure.

Second, the temperature profiles are different. Both our method and that of \citet{2015ApJ...813...99F,2017ApJ...843..150F} 
are dependent on the mid-plane temperature: in \citet{2015ApJ...813...99F,2017ApJ...843..150F} it is through the thermal broadening 
and its connection with the derived non-thermal broadening, while in our simulations it is through the pressure scale height and 
its influence on the depth of the gap created by a planet. \citet{2017ApJ...843..150F} derive a mid-plane temperature that is higher 
than we utilize here, and either reducing the mid-plane temperature used by \citet{2017ApJ...843..150F} 
\footnote{The mid-plane temperature adopted in \citet{2017ApJ...843..150F} is higher than the I16 mid-plane temperature beyond 
$\sim$ 240 au (Figure \ref{fig:scaleheight}). Given a line width, if the thermal broadening term (temperature) decreases, the 
contribution from non-thermal part (turbulence) has to increase. That being said, the upper limit of $\alpha$ would increase, 
if a colder mid-plane temperature (e.g. I16) was assumed.}, 
or increasing the mid-plane temperature in our simulations, can bring our constraints on the 
$\alpha$-viscosity into closer agreement. Besides, our simulations also show there is some degeneracy of acceptable $\alpha$ values, 
which is not fully characterized in this study. When taking the uncertainties into account, we think the seemingly discrepancy of the 
$\alpha$ in the outer disk between this work and \citet{2017ApJ...843..150F}  is NOT significant.

It is also worth pointing out that $\alpha$-viscosity (i.e. transport of angular momentum) and turbulence are directly
proportional only in the simplistic case in which turbulence is created by dissipation of viscosity on small scales. This might
not be the case if dissipation of angular momentum is operated by, for example, MHD winds like those predicted by non-ideal MHD
disk models \citep{2016ApJ...821...80B}.  In this case, the inward radial motion of the gas would be mostly laminar and the
overall turbulence will be low. \citet{2017arXiv171104770S} point out that even with a magnetic field strong enough to 
create a wind and a laminar flow (e.g. the right side of their Figure 2), there are still pockets of weaker magnetic field that 
generate substantial turbulence. This suggests that having a wind (and its associated high-alpha) doesn't necessarily mean that 
the turbulent gas motion will be small. While winds may create the high $\alpha$ and low-turbulence conditions required of the 
models presented here, more work is needed to understand if these are the exact conditions of a magneto-thermal wind. 
Future observations of molecular lines of ring-like disks such as HD163296, HL Tau, TW Hya,
etc., could possibly help clarifying whether the disk evolution is controlled by laminar or turbulent flows.

\section{Conclusion}

Scenarios such as the grain growth at condensation fronts, zonal flows and Rossby wave instability, are proposed to explain 
rings in continuum emission of protoplanetary disks. Whereas, these theories may not necessarily deplete gas at dust gaps. Tidal
interaction between an embedded protoplanet and the gaseous disk can carve a gas gap \citep{1999ApJ...514..344B}, and 
millimeter-sized grains dust grains naturally concentrate toward a local pressure maximum (edges of gas gaps) due to gas drags. 
So dust is expelled from the gas gap. In the HD 163296 disk, we obtain high resolution dust continuum and CO emissions, both of 
which are depleted in dark rings. The depletion of both gas and dust inside a gap is the hallmark of the planet--disk interaction 
model.

In this study, we present \twod global hydrodynamic simulations with dust and gas to model the observed ringed structures of 
the HD 163296 disk. Using the parametric model introduced by I16 as the fiducial model, we conclude that the ringed structures 
can be explained by the planet--disk interaction scenario. We further perform 3D radiative transfer calculation of dust 
continuum and CO emission. The synthetic emission results are compared with the observations to scrutinize our hydrodynamic modeling. 
Assuming a temperature profile similar to I16, our hydrodynamic modeling suggests that three planets with masses of 0.46, 
0.46 and 0.58 $M_\text{J}$ located at 59, 105 and 160 au can reproduce most
of the observational features. In particular, a planet with 0.46 $M_\text{J}$ in a low effective viscosity ($\alpha \sim 5\times
10^{-5}$) region opens a gap that can be seen from the map of dust continuum while does not manifest itself in CO emission.
The effective $\alpha$ viscosity increases with the radius and levels out at $7.5\times 10^{-3}$ beyond 300 au.

We interpret the variation of the effective viscosity profile as the changes of the disk ionization level. The low viscosity
region within 50 au indicates either a dead zone or a region where MRI is possibly quenched by the Hall effect. The $\alpha$ gradually
increases with the radius because non-ideal MHD effects (the Hall effect and ambipolar diffusion) are phased out with increasing
radius. Recent measurements of disk turbulence through molecular lines are generally in good agreement with our result in 
the inner disk \citep{2016MNRAS.461..385B}, while \citet{2017ApJ...843..150F} claim an upper limit of $\alpha$ viscosity that is 
2.5 times lower than our result in the outer disk. However, the discrepancy can be explained by their assumption of the turbulent 
velocity dispersion being a constant fraction of the local sound speed throughout the disk and a hotter disk mid-plane in their models.

The technique of modeling of gas and dust simultaneously presented here can be applied to other systems with ringed structures 
(such as the HD 169142 disk \citep{2017A&A...600A..72F})
to constrain the planet mass as well as the effective viscosity, which may provide a crucial benchmark to various evolutionary
models of protoplanetary disks.

\acknowledgements 
We thank an anonymous referee for carefully reading our manuscript and many thoughtful comments that greatly improved the 
quality of this paper. A.I. acknowledges support from the NSF Grant No. AST- 1535809. A.I., S.L. and H.L. acknowledge the support
from Center for Space and Earth Sciences at LANL, and H.L. acknowledges the support from LANL/LDRD program. S.J. acknowledges 
support from the National Natural Science Foundation of China (Grant No. 11503092). This paper makes
use of the following ALMA data: ADS/JAO.ALMA\#2013.1.00601.S.  ALMA is a partnership of ESO (representing its member states),
NSF (USA) and NINS (Japan), together with NRC (Canada) and NSC and ASIAA (Taiwan) and KASI (Republic of Korea), in cooperation
with the Republic of Chile. The Joint ALMA Observatory is operated by ESO, AUI/NRAO, and NAOJ. The National Radio Astronomy
Observatory is a facility of the National Science Foundation operated under cooperative agreement by Associated Universities,
Inc.

\textit{Facility:} ALMA

\textit{Software:} LA-COMPASS \citep{2005ApJ...624.1003L,2009ApJ...690L..52L,2014ApJ...795L..39F}, RADMC-3D \citep{Dullemond:2012vq}, radmc3dPy (\url{http://www.ast.cam.ac.uk/~juhasz/radmc3dPyDoc/index.html}). 
\end{CJK*} 
\newpage 
\bibliographystyle{apj} 
\bibliography{apj-jour,references}\label{sec:ref} 
\end{document}